\documentclass[manuscript,screen]{acmart}

\AtBeginDocument{%
  \providecommand\BibTeX{{%
    \normalfont B\kern-0.5em{\scshape i\kern-0.25em b}\kern-0.8em\TeX}}}

\setcopyright{none}
\copyrightyear{2023}
\acmYear{2023}
\acmBooktitle{CHI ’23 Workshop on Designing Technology and Policy Simultaneously: Towards A Research Agenda and New Practice, April 23, 2023, Hamburg, Germany}
\acmDOI{}
\acmConference[Design x Policy @ CHI '23]{CHI Conference on Human Factors in Computing Systems}{April 23, 2023}{Hamburg, Germany}

\begin{document}

\title[Anticipating Unintended Consequences of Technology \\ Using Insights from Creativity Support Tools]{Anticipating Unintended Consequences of Technology Using Insights from Creativity Support Tools}

\author{Rock Yuren Pang}
\thanks{Rock Yuren Pang: ypang2@cs.washington.edu; Katharina Reinecke: reinecke@cs.washington.edu}
\affiliation{%
 \institution{Paul G. Allen School of Computer Science,  University of Washington}
 \city{Seattle}
 \state{Washington}
 \country{USA}}
\email{ypang2@cs.washington.edu}

\author{Katharina Reinecke}
\affiliation{%
 \institution{Paul G. Allen School of Computer Science, University of Washington}
 \city{Seattle}
 \state{Washington}
 \country{USA}}
\email{reinecke@cs.washington.edu}

\makeatletter
\let\@authorsaddresses\@empty
\makeatother

\renewcommand{\shortauthors}{Pang and Reinecke}

\begin{abstract}
  Our society has been increasingly witnessing a number of negative, unintended consequences of digital technologies. 
  While post-hoc policy regulation is crucial in addressing these issues, reasonably anticipating the consequences \emph{before} deploying technology can help mitigate potential harm to society in the first place.
  Yet, the quest to anticipate potential harms can be difficult without seeing digital technologies deployed in the real world.
  In this position paper, we argue that anticipating unintended consequences of technology can be facilitated through creativity-enhancing interventions, such as by building on existing knowledge and insights from diverse stakeholders. Using lessons learned from prior work on creativity-support tools, the HCI community is uniquely equipped to design novel systems that aid in anticipating negative unintended consequences of technology on society.
\end{abstract}



\maketitle

Digital technologies have significantly advanced our life quality but have also had many unintended, negative consequences on society --- outcomes of purposeful actions that are not intended or foreseen at the time of development~\cite{merton1936unanticipated}. 
For example, innovations in interaction design have simultaneously exacerbated the disparities between the experiences of the demographic groups included or omitted from the research and development process~\cite{toyama2015geek}. Research prototypes that were intended to benefit scientists have been met with strong resistance by other researchers, as the models were unable to distinguish truth from falsehood~\cite{heaven_2022}. Accessibility features in operating systems have created security vulnerabilities that can be exploited by attackers~\cite{jang2014a11y}.

Our recent work has shown that many technologists want to anticipate potential unintended consequences of their research innovations on society but that they lack the know-how and specific approaches for doing so~\cite{important}. The observation is concerning and somewhat surprising given the increasing discussions of ethics in the news and computer science curricula, as well as proposed changes in the peer review systems including impact statement in HCI and AI~\cite{hecht2021s, IUI2022, ACL2022, NeurIps2021}. 
While requiring comprehensive ethics training is a vital step to make technologists fully aware of the societal implications, achieving this is likely a long-term process that may require years of experience~\cite{important}. We need to support developers to more easily anticipate  various outcomes of technology, but reasonably doing so for certain user groups, non-user stakeholders, or society as a whole is a difficult task rife with uncertainties~\cite{nanayakkara2020anticipatory}.

To bridge this gap, this paper proposes to enhance related HCI frameworks that account for human values and social impact during design
by combining insights from the creativity support tool literature. After all, anticipating unintended consequences from various perspectives is inherently a creative process.
We encourage researchers and policymakers to collaboratively envision methods and tools to anticipate adverse societal effects in this workshop. 

\emph{Prior Work in HCI} have proposed frameworks --- such as the value sensitive design (VSD)~\cite{friedman2008value, hendry2021value} and design fictions~\cite{bleecker2022design} --- to guide anticipating potential unintended consequences.
VSD starts with the motivation of understanding the technology, human value, or context of use. Then, it identifies direct and indirect stakeholders and the benefits and harms to each stakeholder. Benefits and harms are mapped to the corresponding values, and value conflicts are identified. VSD aims to help practitioners find alternative approaches that uphold their chosen values while accommodating the same constraints. 
Another relevant approach for considering unintended consequences is design fiction~\cite{bleecker2022design, Baumer2018WhatWY}, a form of speculative design~\cite{Lindley2016PushingTL, Dunne2013SpeculativeED}. It envisions new futures including both the technical aspects and the social and political outcomes of a world with that technology~\cite{lindley2017implications}.
To support these approaches in practice, recent toolkits have been developed for designers~\cite{chivukula2021surveying}, such as the Envisioning Cards~\cite{EnvisioningCards}, Metaphor Cards~\cite{MetaphorCards}, and Design Fiction Memos~\cite{wong2021using}. They typically offer concrete guidance such as thought-provoking questions (e.g., one question in the Envisioning Cards is ``what challenges will be encountered by your system if it is used in other countries?'').

While the frameworks and tools guide developers to consider societal implications, they are not without limitations. For example, the toolkits often target practitioners and were used in group sessions~\cite{gray2019ethical}, so applying them to research projects may pose additional challenges. Design fiction may suffer from exaggeration~\cite{malpass2013between} and be difficult to evaluate~\cite{Baumer2018WhatWY}. These approaches to consider societal implications were often assumed to be effective in practice~\cite{gray2019ethical}, but adopting them can be difficult, especially for most developers who are not familiar with these concepts. Using the current tools assumed users have sufficient knowledge needed to anticipate the consequences. In fact, our recent interview with 20 researchers found that no participant routinely used these frameworks or tools in their own research~\cite{important}. 

\emph{Creativity Support Tools} literature, on the other hand, introduced concepts and tools and demonstrated that properly scaffolding can galvanize users' creativity~\cite{macneil2021framing}, thus improving the practicality of an abstract or unfamiliar process. Scaffolding refers to structure and guidance to ensure successful completion of a task~\cite{wood1976role}, using approaches such as rich examples~\cite{siangliulue15:providing, Ngoon2018InteractiveGT}, expert patterns~\cite{kim2015motif}, instructions~\cite{pandey2018docent}, and templates~\cite{bauer2013designlibs, aperitif}. For example, Siangliulue et al. found that a collection of timely examples can support design ideation~\cite{siangliulue15:providing}. Sampling diverse inspirational examples and providing a visual overview of the ideas have been shown to improve people's brainstorming activity~\cite{siangliulue16:ideahound-uist}. Similar work on cognition and creativity support confirmed that examples inspire and unveil new ideas most often in the design contexts such as design images and writing feedbacks~\cite{Ngoon2021ShwnAC, Kang2018ParagonAO, Ngoon2018InteractiveGT, DiFede2022TheIM}. 
 To organize these examples, prior work has used categories of certain topics and characteristics, which were proved essential to human cognition~\cite{Rosch1976BasicOI, Ward1994StructuredIT}, to facilitate exploration of the multitude of examples.
 For example, IdeaRelate facilitates the exploration of examples in the context of COVID-related topic ideation by tagging them into different topics, helping users to include more perspectives in their own idea generation~\cite{idearelate}. 

 To enhance the current frameworks and toolkits, the HCI community can leverage the insights from the creativity literature to design and evaluate future support for technology developers. One concrete direction is to curate examples from past incidents and cluster them in an idea or a solution space to support the generation of relevant ideas for developers, an idea similar to~\cite{siangliulue16:ideahound-uist}. These examples can be obtained from online articles~\cite{neumann2008risks} and experts~\cite{kim2015motif}.  An important challenge is to collect insights from a diverse population to avoid missing voices from non-WEIRD communities~\cite{sebastian, diverse}.
 Policymakers can additionally provide specific policy changes following a previous introduction of technology and establish expert patterns~\cite{kim2015motif} to consider the societal impact for technologists to follow. We believe that this workshop can enrich the discussion on how to proactively anticipate unintended consequences routinely in the broader research community.

\bibliographystyle{ACM-Reference-Format}
\bibliography{bibliography}

\end{document}